# Signature of topological band crossing in ferromagnetic $Cr_{1/3}NbSe_2$ epitaxial thin film

Bruno Kenichi Saika,[1] Satoshi Hamao,[1] Yuki Majima,[1] Xiang Huang,[1] Hideki Matsuoka,[2] Satoshi Yoshida,[1] Miho Kitamura,[3] Masato Sakano,[1,4] Tatsuto Hatanaka,[1] Takuya Nomoto,[5] Motoaki Hirayama,[1,2,4] Koji Horiba,[3] Hiroshi Kumigashira,[6] Ryotaro Arita,[5,2] Yoshihiro Iwasa,[1,2,4] Masaki Nakano,[1,2,4] and Kyoko Ishizaka[1,2,4]

[1]*Department of Applied Physics, University of Tokyo, Tokyo 113-8654, Japan*

[2]*RIKEN Center for Emergent Matter Science (CEMS), Wako 351-0198, Japan*

[3]*Photon Factory, Institute of Materials Structure Science,*

*High Energy Accelerator Research Organization, Tsukuba 305-0801, Japan*

[4]*Quantum-Phase Electronics Center (QPEC),*

The University of Tokyo, *Wako 113-8656, Japan*

[5]*Research Center for Advanced Science and Technology,*

The University of Tokyo, Komaba, Meguru-ku, Tokyo *153-8904, Japan*

[6]Institute of Multidisciplinary Research for Advanced Materials (IMRAM),

Tohoku University, Sendai 980–8577, Japan

(Dated: September 15, 2022)

# Abstract

In intercalated transition metal dichalcogenides (I-TMDC), transition metal intercalation introduces magnetic phases which in some cases induce topological band crossing. However, evidence of the topological properties remains elusive in such materials. Here we employ angle-resolved photoemission spectroscopy to reveal the band structure of epitaxially grown ferromagnetic $Cr_{1/3}NbSe_2$. Experimental evidence of the Weyl crossing shows $Cr_{1/3}NbSe_2$ to be a topological ferromagnet. This work highlights I-TMDC as platform towards the interplay of magnetic and topological physics in low-dimensional systems.

Nontrivial topology in the band structure of solids can have profound effects to the macroscopic properties of matter, with potential technological applications ranging from dissipationless transport to many exotic quasiparticle excitations [1–3]. Weyl semimetals refer to one of such topological phases, in which broken spatial-inversion or time-reversal symmetry gives rise to pairs of singularity points of source (positive chirality) or sink (negative chirality) of the Berry curvature [4–7]. To date, majority of Weyl semimetal have been found in materials with broken inversion symmetry, such as TaAs [8, 9], $PbTaSe_2$ [10, 11], $TaIrTe_4$ [12, 13] and $MoTe_2$ [14, 15]. In addition, time-reversal symmetry breaking Weyl semimetals have also been found in ferromagnetic compounds such as $Co_2MnGa$ [16, 17] and $Co_3Sn_2S_2$ [18, 19]. However, magnetic Weyl semimetals are still restricted to a handful class of materials, and several predicted systems still lack experimental evidence. Expanding the family of time-reversal symmetry breaking Weyl systems is of extreme relevance, as varied magnetic textures are expected to couple with topologically non-trivial electronic structure to serve as platform for emergent quantum phases. Additionally, novel routes towards accessing and controlling topological phases via external fields are expected in such coupled systems.

The 3$d$-transition metal intercalated transition metal dichalcogenides (3$d$ I-TMDC) represent a class of intercalated layered materials which possess rich magnetic phases due to the localized magnetic moment of intercalants [20–22]. Recently, the magnetic 3d I-TMDC have been re-examined under the framework of topological physics. The antiferromagnet $Co_{1/3}NbS_2$ has attracted considerable attention as potential non-collinear spin textures and topological band crossings were linked to the giant anomalous Hall effect observed in the compound [23–25]. Additionally, first-principles calculation focused on the ferromagnetic 1/3-intercalated $M_{1/3}TX_2$ ($M$: 3d-transition metal; $T$: Nb, Ta; $X$: S, Se) has suggested the presence of band crossing points in the electronic structure, indicating such systems to be candidates for magnetic Weyl materials [26]. However, several difficulties hinder the progress in these materials such as the lack of easy cleavage planes, and surface disorder effect [27]. This stems from the nature of the intercalation itself, as occupancy of intercalant atoms at the cleaved surface produces uncertainty of terminations. In addition, the study of the topological phases in I-TMDC is complicated by complex magnetic ordering which provides a non-trivial modification to the electronic structure. Hence, clear demonstration of such topologically non-trivial ferromagnetic phases requires 1/3-intercalated systems with accessible magnetic phases. From the consolidated ferromagnetic $M_{1/3}TX_2$ candidates, the $Cr_{1/3}NbS_2$ and $Mn_{1/3}NbS_2$ exhibit complicated helimagnetic ordering, whereas the selenide counterpart $Cr_{1/3}NbSe_2$ exhibits a rather simpler easy-plane ferromagnetic phase [28,29]. Consequently, $Cr_{1/3}NbSe_2$ provides a suitable candidate for visualizing the newly predicted topological phases in $3d$ I-TMDC.

In this Letter, motivated by the prospects of magnetic topological phases in $3d$ I-TMDC, we provide new evidence to the topologically non-trivial band structure of the ferromagnetic $M_{1/3}TX_2$ by targeting $Cr_{1/3}NbSe_2$. We report the electronic structure of $Cr_{1/3}NbSe_2$ fabricated via molecular-beam epitaxy (MBE) and measured via angle-resolved photoemission spectroscopy (ARPES). We observe the correspondence between the electronic structure of the multi-layer $Cr_{1/3}NbSe_2$ epitaxial thin-films and the theoretical ferromagnetic bulk band structure. Through careful examination of the photoemission spectra, we find evidence of the Weyl crossing predicted in first-principles calculation.

The 9-layer $Cr_{1/3}NbSe_2$ thin-films were fabricated onto atomically flat sapphire substrates using MBE, in which the number of layers is defined from the number of intercalant layers. Niobium was evaporated via electron-beam evaporator, whereas chromium and selenium were evaporated by standard Knudsen cells. The samples were grown and annealed at 850 C under constant Se flux. Thick Se cap was deposited before exposure to the atmosphere for ex-situ measurements. The growth procedure was monitored via reflection high-energy electron diffraction (RHEED). Prior to ARPES and low-energy electron diffraction (LEED) measurements, samples were annealed at temperatures above 170°C in ultrahigh vacuum for cap removal. X-ray photoemission spectroscopy (XPS) and ARPES measurements were performed at 20 K in the photoelectron spectroscopy end-station equipped with Scienta-Omicron SES2002 electron analyzer at BL-2A (MUSASHI) in Photon Factory, Japan. Vacuum ultraviolet ARPES (VUV-ARPES) and soft x-ray ARPES (SX-ARPES) were conducted with photon energies ranging from 80 to 150 eV (circular polarization) and 250 to 400 eV (linear polarization), respectively. Energy resolution was estimated to be 75 meV for VUV-ARPES, and 130 meV for SX-ARPES. The calculated band structure was obtained using the projector augmented wave (PAW) with the Perdew-Burke-Ernzerhof (PBE) exchange-correlation functional, as implemented in Vienna ab initio simulation package (VASP) [30–33]. Cut-off energy and *k*-points mesh were set to 500 eV and 12×12×6, respectively.

The bulk crystal structure of the $Cr_{1/3}NbSe_2$ is shown in Figs. 1(a,b). $NbSe_2$ layers are composed of Nb atoms surrounded by six Se atoms in the trigonal prismatic coordination. The layers follow the $2H_a$ stacking order which creates octahedral sites at the van der Waals gaps directly above and below Nb atoms. As shown by the arrows in Figs. 1(a,b), Cr atoms occupy these 2c-Wyckoff sites and construct a non-centrosymmetric structure described by the space group $P6_322$ [20, 21, 34, 35]. Cr atoms ordering creates a (√3×√3)R30° superstructure unit cell relative to the parent $NbSe_2$, which translates into a smaller hexagonal Brillouin zone (BZ) rotated 30° relative to the parent $NbSe_2$ BZ as shown in Fig. 1(c). To differentiate from the parent BZ, the high-symmetry points of the $Cr_{1/3}NbSe_2$ BZ are denoted with the prime symbol (*M'*, *K'*, *H'* and *L'*). Fig. 1(d) and (e) show the calculated bulk band structures along *M*-Γ-*K* in the paramagnetic and ferromagnetic phases, respectively. As can be seen, the electronic bands near the Fermi level are strongly modified by introducing the magnetism. Notably, a band crossing close to the Fermi level is observed to occur along the Γ-M (Γ-*K'*) direction in the ferromagnetic phase, originating from the spin-up components. According to calculation, this band crossing forms a Weyl point (WP), and the bands participating in the Weyl crossing correspond to the niobium-like 4*d* and the chromium-like 3*d* derived states. The Weyl points and the corresponding chirality are shown schematically relative to the original $NbSe_2$ BZ in Fig. 1(f). We also compared the calculated ferromagnetic band structure in the presence of spin-orbit coupling (SOC), in which the overall electronic structure is not significantly altered and the band crossing is still observed [36].

The characterization of the $Cr_{1/3}NbSe_2$ epitaxial thin-films is described in Fig. 2. The superstructure produced by Cr intercalation was confirmed via low-energy electron diffraction (LEED) in Fig. 2(a). The ($\sqrt{3} \times \sqrt{3}$)R30° diffraction spots are clearly observed in the interior of the parent NbSe$_2$ diffraction pattern, indicating the 1/3 intercalation ratio of Cr-atoms. The out-of-plane x-ray diffraction (XRD) taken along the [001] direction is shown in Fig. 2(b). The presence of the strong diffraction peak at $2\theta$ = 14° along with the Laue oscillation indicate the high crystalline quality of the film along the out-of-plane direction. The reported number of layers was obtained from Laue oscillation period. We also performed x-ray photoemission spectroscopy (XPS) and observed the clear presence of the Nb, Se and Cr core level spectra [36]. The magnetization as function of temperature (M-T) measured in the field-cooling (FC) condition is displayed in Fig. 2(c), in which the magnetic field of 100 Oe was applied along the in-plane direction (H⊥$c$). The M-T curve clearly shows the presence of a magnetic phase transition at around 66 K, as defined from the maximum of the M-T curve derivative relative to temperature. The magnetic field dependence of the magnetization measured at 2 K is displayed in the inset of Fig. 2(c). The near zero-field magnetization is also shown in the right inset, in which a finite spontaneous magnetization can be observed. The observed easy-plane ferromagnet character agrees with the properties found in the bulk polycrystalline Cr$_{1/3}$NbSe$_2$[28, 29, 37].

Due to the few-layer nature of the Cr$_{1/3}$NbSe$_2$ thin-film samples, it is non-trivial whether the band structure can be expected to exhibit bulk-like electronic structure predicted to host topological phases. To resolve this question, the overall three-dimensional electronic structure of the multi-layer Cr$_{1/3}$NbSe$_2$ was obtained via SX-ARPES, and the results are displayed in Fig. 3. The observed Fermi surface (FS) of Cr$_{1/3}$NbSe$_2$ is characterized mainly by the presence of two rather isotropic hole-bands centred around $\bar{\Gamma}$ and $\bar{K}$, and one 'flower-like' shape around $\bar{\Gamma}$ as shown in Figs. 3(a,b). In Fig.3 (a), the smaller (larger) FS observed around $\bar{\Gamma}$ is labeled as $\alpha$-band ($\beta$-band), whereas the 'flower-like' FS is denoted as $\gamma$-band. The respective in-plane dispersion of the $\alpha$- and $\beta$-bands together with the guide-for-eyes traces are displayed in the Fig. 3(b). The Fermi momentum ($k_F$) along the $\bar{\Gamma}$-$\bar{M}$ direction ($k_x$) of the inner and outer bands were estimated to be roughly 0.16 Å$^{-1}$ and 0.34 Å$^{-1}$, respectively. We also display the band unfolding calculation considering the ferromagnetic ground-state without the inclusion of spin-orbit coupling (SOC) in Fig. 3(c). The minority and majority-spin bands are depicted with blue and red markers, respectively. From the calculation results, it is possible to see the presence of two hole-like bands with strong weight centred around $\bar{\Gamma}$, respectively of spin-down (inner) and spin-up (outer), which agrees well with the measured $α$- and $β$-bands in Fig. 3(b). It is worth mentioning that such agreement is lost when compared to the paramagnetic band structure displayed in Fig. 1(d), particularly regarding $β$-bands which are absent in the corresponding energy-momentum of the calculation.

ARPES experiments were further conducted by varying photon energies from 250 to 382 eV to characterize the electronic structure along the out-of-plane direction ($k_z$). The estimated value for the inner potential is 15.7 eV. Fig. 3(d) schematically shows the measurement range. The FS measured along $k_x$-$k_z$ plane is displayed in Fig. 3(e), in which the solid red curve displays the $k_z$ vs $k_x$ cut taken with 363 eV shown in Fig. 3(a,b). Notably, $α$- and $β$-bands exhibit different degrees of dimensionality along $k_z$, namely $α$-bands exhibit a 3D-like spheroidal FS, whereas $β$-bands exhibit a cylindrical FS. The dispersion of the $α$-bands along the $k_z$ direction is shown in Fig. 3(f). The guide-for-eyes trace the main dispersive features of the $α$-bands, showing the strong three-dimensionality. The ARPES images without guide-for-eyes are given in [36]. Fig. 3(g) shows the band-unfolding calculation results along the Γ-$A$ direction. The presence of down-spin bands with strong $k_z$ dispersion can be clearly visualized, being consistent with the previous claim regarding the correspondence between $α$- ($β$-) and down-spin bands (up-spin bands). This result suggests an electronic structure of multi-layer $Cr_{1/3}NbSe_2$ close to bulk. Indeed, the bulk band calculation is in good correspondence with ARPES, particularly when taking into account the 4π $k_z$-periodicity of the photoemission spectra. Considering the $k_z$ dispersion, it is notable that the photoemission intensity of the $α$ bands exhibits an apparent twofold period, repeating itself at every odd Γ point. Such 4π-period has been reported in other materials with nonsymmorphic space group, such as $T_d$-$MoTe_2$ [38], $T_d$-$WTe_2$ [39], $WSe_2$ [40] and graphite [41]. Once such effects are taken into consideration, the agreement between the bulk and $Cr_{1/3}NbSe_2$ thin-films electronic structure becomes more apparent.

Now we examine the potential topological features of the bands obtained using SX- and VUV-ARPES. The low-energy features measured with 363 eV photons along $k_x$ are displayed in Fig. 4(a). Although the previously mentioned $β$-bands are strongly apparent, we can also discern the presence of additional intensity very close to $E_F$. By examining the momentum distribution curves (MDC), we can note the presence of an additional peak at roughly 0.57Å$^{-1}$ at $E_B$ = 0.0 eV which shifts towards smaller momentum values at higher binding energies as shown by the green arrows in Fig. 4(b). The peaks eventually fade due to the strong intensity of the Nb-derived $\beta$-bands burying the signal, indicating the crossing point around $E_B$ = 0.1 eV. The feature resembles the dispersion of the $\gamma$-bands described in Fig. 3(c). The calculated FS is shown in Fig. 4(c). The Cr-derived $\gamma$-bands are observed to construct a 'flower-like' FS centred around Γ with the 'petals' aligned along the Γ-M direction. To experimentally visualize the evolution of the $\gamma$–bands as it approaches the crossing-point, the constant energy maps integrated over a 40 meV window are displayed in Fig. 4(d)-(f). At the Fermi-level (Fig. 4(d)), the 'flower-like' FS analogous to the calculation is clearly discernible, which shrinks with increasing $E_B$ and eventually disappears at $E_B$ = 120 meV.

Additional supporting evidence to the topological features was obtained via high-resolution VUV-ARPES measurements. Considering the photoionization of each element, the Cr 3$d$ cross-section relative to Nb 4$d$ and Se 4$p$ cross-sections increase from SX to VUV region [42]. The spectra measured along $k_x$ with 84.5 eV photons are displayed in Fig. 4 (g). To enhance the visibility of the band crossing, filtering of the ARPES image was performed by convolution of a 2D-Gaussian band pass filter following the Fourier space approach in Fig. 4 (h) [43]. Energy distribution curves (EDC) also provide visual guide to the observation of the weak but discernible signal of the band crossing in Fig. 4(i). Notably, both SX- and VUV-ARPES indicate the band crossing occurring at roughly $E_B$ = 0.1 eV and momentum $k_x$ = 0.40 Å$^{-1}$.

The topological nature of the observed band crossing can be understood from the twofold rotation symmetry ($C_2$) along the 100 axis in real space. This creates six Weyl nodes along Γ-M. The situation is analogous to the previously reported theoretical works on V$_{1/3}$NbS$_2$ and Mn$_{1/3}$NbS$_2$ [26]. In fact, the presence of the topological features is not restricted to the Cr$_{1/3}$NbSe$_2$ system, and similar Weyl nodes are expected to emerge with different intercalant atoms and host layers. Considering the wide variety of magnetic phases promoted by intercalation, 3$d$ I-TMDC provide the ideal platform to realize non-trivial topological systems with complex magnetic structures. Additionally, the layered nature of the 3$d$ I-TMDC provides further prospects towards novel two-dimensional systems, as further tailoring of the magnetic and topological physics can be achieved in the three- to two-dimension crossover. In this context, not only we observed such topological features, but we also demonstrate its feasibility in epitaxial thin-films fabricated via MBE. Such thin-film approach to intercalated systems could readily expand the access to novel magnetic Weyl semimetals, while providing routes towards the functionalization in device applications. Novel prospects towards dimensionality control in magnetic systems as well as heterointerfaces could provide exciting and yet unexplored perspectives in epitaxially grown $3d$ I-TMDC.

In conclusion, by employing MBE, we successfully examined the electronic structure of epitaxially grown $Cr_{1/3}NbSe_2$. ARPES showed the three-dimensional character of the electronic structure, and comparison with bulk band calculation indicated the correspondence between the $Cr_{1/3}NbSe_2$ thin-films and bulk. We have identified the presence of a weak signal of the predicted band crossing, further evidencing $Cr_{1/3}NbSe_2$ as potential ferromagnetic topological material. The present results provide proof-of-concept evidence of the MBE-grown $3d$-intercalated TMDC as platform towards novel magnetic Weyl systems.

**ACKNOWLEDGEMENTS**

B.K.S. acknowledges the support by the World-leading Innovative Graduate Study Program for Materials Research, Information, and Technology (MERIT-WINGS) of the University of Tokyo. The measurements were performed under KEK-PF (Proposal No. 2020G634). This work was partly supported by the JSPS KAKENHI (No. JP19H05826, No. JP20H0183, No. JP21H05235, No. JP22H00107), CREST JST (No. JPMJCR20B4) and PRESTO JST (No. JPMJPR20L7).

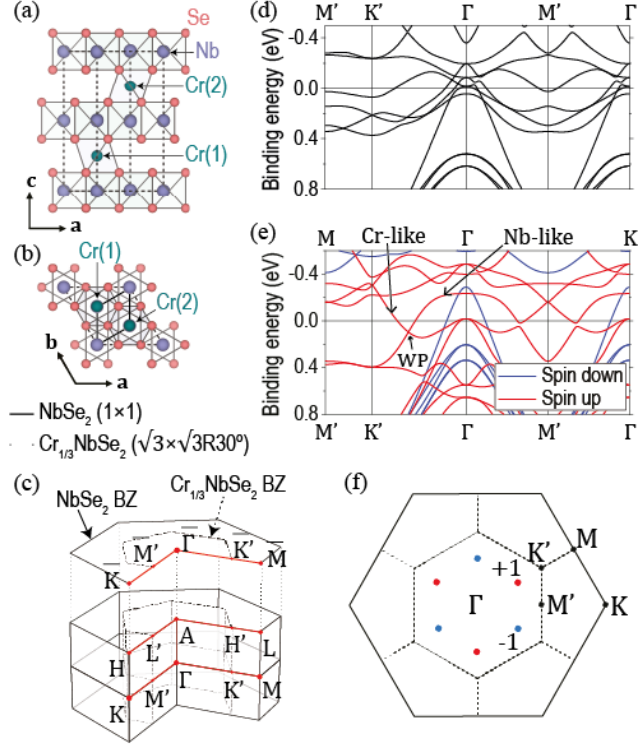

FIG. 1. Crystal structure and band calculation of $Cr_{1/3}NbSe_2$. Crystal structure viewed along *b*-axis (a) and *c*-axis (b). Solid and dotted lines denote the $NbSe_2(1 \times 1)$ and $Cr_{1/3}NbSe_2$ ($\sqrt{3}\times\sqrt{3}$)R30° cell, respectively. (b) Position of Cr(1) and Cr(2) atoms highlighted over the top-view of $Cr_{1/3}NbSe_2$. (c) Hexagonal BZ and high-symmetry points of the $NbSe_2$ (solid lines) and $Cr_{1/3}NbSe_2$ (dotted lines). Bulk band structure calculation in the (d) paramagnetic and (e) ferromagnetic states. Nb-like and Cr-like band crossing can be observed along Γ-M in the ferromagnetic band structure. (f) Schematic position of the WP. Red and blue colors denote the positive and negative chirality WP, respectively.

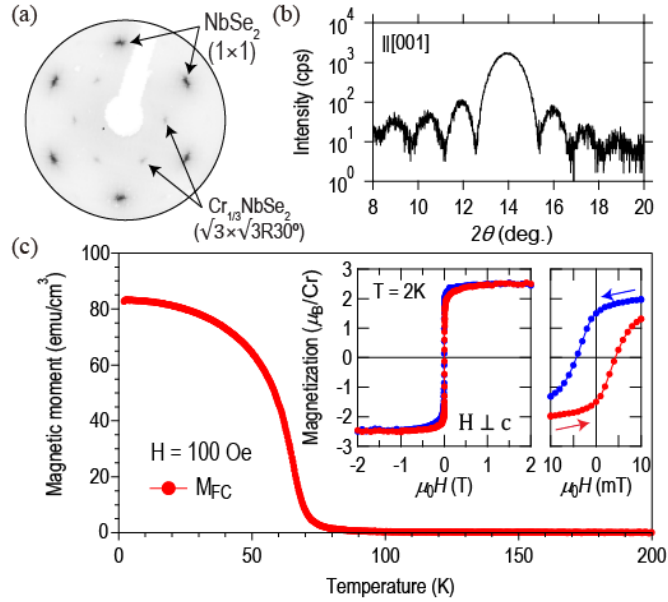

FIG. 2. $Cr_{1/3}NbSe_2$ thin-films characterization. (a) LEED pattern of $Cr_{1/3}NbSe_2$ taken at room temperature showing the presence of (√3×√3)R30° diffraction spots. (b) Out-of-plane XRD pattern exhibiting Laue oscillation period. (c) Temperature dependence of the magnetic moment of $Cr_{1/3}NbSe_2$ thin-film. Magnetic field of 100 Oe was set parallel along the $NbSe_2$ layers (H⊥c). Inset left: in-plane magnetic field (H⊥c) dependence of the magnetization taken at 2 K (blue and red markers indicate the direction of the decreasing- and increasing-field sweep, respectively). Inset right: close-up of the near zero-field magnetization.

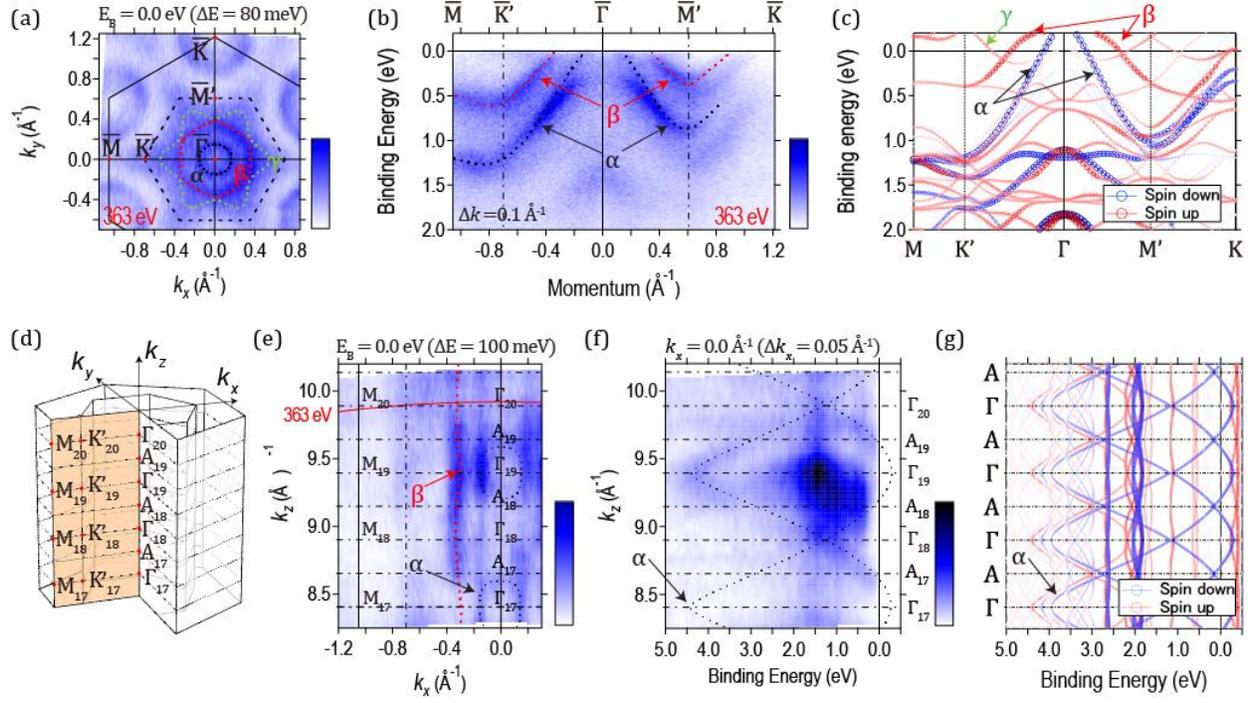

FIG. 3. SX-ARPES characterization of electronic structure. (a) FS mapping taken with 363 eV excitation energy and integral window of 80 meV. Guide-for-eyes trace bands around Γ with labels $\alpha$ (black), $\beta$ (red) and $\gamma$ (green), respectively. (b) Dispersion taken with momentum windows of 0.1 Å$^{-1}$ along the 2D high-symmetry points $\bar{M}$-$\bar{\Gamma}$-$\bar{K}$ (c) Band unfolding onto the original NbSe$_2$ BZ in the ferromagnetic state with blue (red) markers representing spin-down(up) bands. Arrows show corresponding $\alpha$-, $\beta$- and $\gamma$-bands. (d) Schematic diagram of the measurement range along $k_z$. (e) Constant-energy mapping ($E_B$ = 0 eV and integral window $\Delta E_B$ = 100 meV) taken along $k_z$. Red full-line shows the cut corresponding to the 363 eV measurement shown in (a) and (b). Guide-for-eyes highlights the $\alpha$- and $\beta$-bands along $k_z$. (f) Band dispersion along Γ-A (integral window $\Delta k_x$ = 0.1 Å$^{-1}$). Guide-for-eyes trace the visible $\alpha$-bands. (g) Band unfolding calculation along Γ-A.

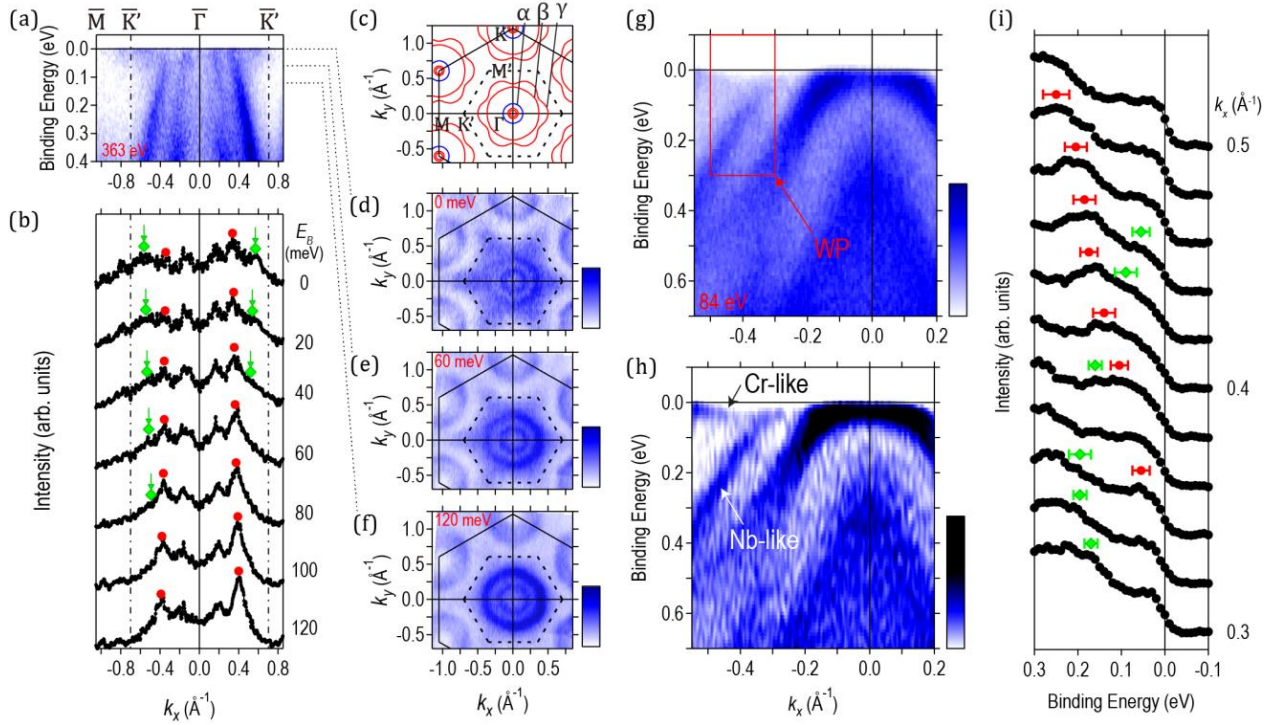

FIG. 4. Evidence of band crossing in the SX- and VUV-ARPES. (a) In-plane dispersion along $\bar{\Gamma}$-$\bar{M}$ taken with 363 eV photon energy. (b) Momentum distribution curves (integration width: 0.02 eV) along $\bar{\Gamma}$-$\bar{M}$. Green diamond (red circle) markers display bands with inward (outward) dispersion. (c) Calculated FS of $Cr_{1/3}NbSe_2$ showing corresponding $\alpha$-, $\beta$- and $\gamma$-bands. (d)-(f) Constant energy maps taken at different binding energies showing the change to the 'flower-like' Fermi surface (integration width: 0.04 eV). (g) High-resolution VUV-ARPES measured with 84.5 eV photon energy along the $\bar{\Gamma}$-$\bar{M}$. (h) Band filtering of the spectra in (g) using convolution of a 2D Gaussian filter. Arrows indicate bands participating in the crossing. (i) Energy distribution curves (integration width: 0.02 Å$^{-1}$). Peak-plot traces Cr-like (green diamond) and Nb-like (red circle) bands.